# DECISION TREES FOR COMPLEXITY REDUCTION IN VIDEO COMPRESSION

*Natasha Westland, André Seixas Dias, Marta Mrak*

British Broadcasting Corporation, Research & Development Department
London and Salford, UK

## ABSTRACT

This paper proposes a method for complexity reduction in practical video encoders using multiple decision tree classifiers. The method is demonstrated for the fast implementation of the 'High Efficiency Video Coding' (HEVC) standard, chosen because of its high bit rate reduction capability but large complexity overhead. Optimal partitioning of each video frame into coding units (CUs) is the main source of complexity as a vast number of combinations are tested. The decision tree models were trained to identify when the CU testing process, a time-consuming Lagrangian optimisation, can be skipped i.e a high probability that the CU can remain whole. A novel approach to finding the simplest and most effective decision tree model called 'manual pruning' is described. Implementing the skip criteria reduced the average encoding time by 42.1% for a Bjøntegaard Delta rate detriment of 0.7%, for 17 standard test sequences in a range of resolutions and quantisation parameters.

## 1. INTRODUCTION

As the popularity of higher definition video increases, so does the need for more bit rate efficient compression techniques, especially for live use cases. To handle the growing requirements, research is ongoing to create algorithms capable of representing video sequences in as little information as possible. More recent video coding standards, including the Versatile Video Coding (VVC) currently under development, incorporate sophisticated prediction techniques and flexible partitioning structures in block-based hybrid coding to achieve this. This approach enables efficient reduction of bit rate, which has already been proven in High Efficiency Video Coding (H.265 / HEVC) and the recent AV-1 video coding standards [2].

However, the combination of tools used to achieve a higher coding efficiency results in high computational complexity at the encoder side [3]. This property of modern video codecs needs to be moderated in practical implementations, which often significantly sacrifice compression efficiency to achieve a given complexity reduction target. Contrary to other research in this area, discussed further in Section 2, this work focusses on a practical encoder rather than a test model. This gives a better understanding of the tangible benefits from an optimisation in practice. A simpler machine learning technique is preferred over a complex one, both temporally and computationally, as the practicality of the solution is prioritised.

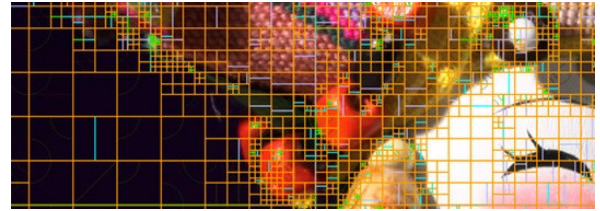

**Fig. 1**. A coding unit partitioning example for a sequence portion.

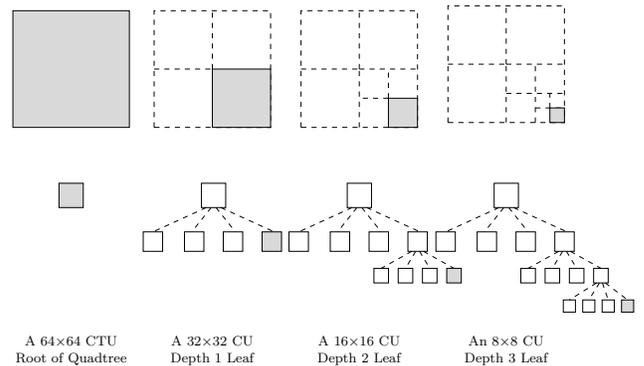

A 64×64 CTU  A 32×32 CU  A 16×16 CU  An 8×8 CU
Root of Quadtree  Depth 1 Leaf  Depth 2 Leaf  Depth 3 Leaf

**Fig. 2**. A schematic showing the relative sizes of CUs at different 'depths', and the corresponding quad-tree partition structure.

The method proposed in this paper reduces the complexity of a fast encoder implementation by targeting a particularly computationally intensive process within the encoder. In HEVC, a single video frame is split up into smaller units to minimise the number of bits needed to represent its information. For example, a section with no detail and motion (from one frame to the next) such as a stationary area of black background needs relatively few bits to describe it, and subsequently can be encoded in a large block, as demonstrated in Figure 1. The inverse is true for fast moving and high detail areas.

The constituent blocks of spatial video are called Coding Units (CUs) and in HEVC can range in size from 64×64 to 8×8 pixels, combinations of which form a flexible quad-tree structure as shown in Figure 2. To determine if a CU should be split up, Rate Distortion Optimisation (RDO) is implemented as a brute force approach to optimise the Lagrange function:

$$\min[J(m)] \text{ where } J(m) = D(m) + \lambda \cdot R(m), \quad (1)$$

where m indicates a combination of CU parameters, $R$ is the rate, $D$ is the distortion (sum of squared differences), $J$ is the Rate-Distortion (RD) cost and $\lambda$ is the Lagrange multiplier. For a $64 \times 64$ top level CU there are > 80,000 possible values for $m$. $R$

and $D$ are examined $\forall$ $m$ until the optimal values are found which minimise $J$ for a given $\lambda$.

If a metric can be discerned by which, if satisfied, the RDO process can be terminated early or skipped altogether, then the complexity of the algorithm is reduced. This metric should be able to identify the likelihood that a given CU can stay whole and all $m$ combinations do not need to be checked. If all CU splitting depths could be correctly predicted, then the encoding process could be 75% less complex [4]. This claim formed the starting point of the research described in this paper.

In the proposed approach, features related to each CU are collected at the point in encoding just before the computationally expensive RDO takes place. Using these, three decision tree models were trained (offline) one for each CU depth at which a decision takes place. The models could identify the most probable feature combination which, when used as the decision criterion, the RDO could be terminated early or skipped if possible. The proposed optimisations were applied on top of an efficient implementation of an HEVC encoder [5], in order to show the relevance of the proposed work to practical video encoding implementations. Such fast implementations do not exist for the VVC standard under development, hence an HEVC test environment was used.

A brief background on previous research in the field and Decision Tree theory can be found in Section 2. More detail about the proposed method and 'manual pruning' is discussed in Section 3. The experimentation and development are outlined in Section 4 and a discussion of results can be found in Section 5.

## 2. BACKGROUND

Previous work has been done to reduce the complexity of the HEVC test model (HM) which is the reference codec software for the latest standard. In 2015, Y. Zhang et al. used a method of several support vector machines (SVMs) to reduce the complexity of the CU splitting process [4]. Their method managed to speed up HM, which has little optimisation by default, by an average of 52% for an additional bit rate of approx. 2%. A similar study by Heindel et al. also used SVMs to tackle this problem with respect to HM, reducing the encoding time by an average of 60% for a bit rate increase of 3-4% [6]. The success of these studies motivated the research into this area with respect to faster HEVC encoder implementations.

Inspired by this, SVMs were initially researched, but the importance of keeping the skipping method simple was highlighted during the experimentation phase. An average encoding time decrease of 12% was achieved for a model using 7 features. However, the process of performing the decision equation was often time consuming itself, resulting in varied results which were largely dependent on video content.

The 'decision function' found by the trained SVM model is a hyperplane equation, which, when given a vector of feature values, returns a probability that the CU in question can be kept whole. In an SVM, especially for a high dimensional feature spaces, the boundary equation can become complex itself if not properly optimised, as it involves large sums of vector multiplication. Therefore, the novel part of the proposed method is to use 'manually pruned' decision trees instead of SVMs for the decision criteria. These are extremely quick and simple to implement as they merely filter the coding units based on a few feature values. Earlier tests of the proposed method with the practical HEVC encoder implementation, saved on average 30.2% more time than the SVM with the same sequences.

### 2.1. Decision Trees

The Classification and Regression Tree (CART) algorithm was developed in 1984 by Breiman et al [7]. It can be used to build classification trees such as the ones used in this research. The algorithm can determine the optimal cut criteria by using a simple metric called the Gini Impurity (GI) recursively at each decision node, described by:

$$GI(t) = 1 - \sum_{k=1}^{K} p_k^2 \qquad (2)$$

where $K$ is the total number of classes, $k$ is a given class and $p_k$ is the probability of the sample belonging to class $k$, i.e:

$$p_k = \frac{\text{no. samples in class } k}{\text{total samples available at current node, } t}. \qquad (3)$$

Eq. 2 is evaluated at each 'node' of the tree. The maximum GI is $(1 - 1 / K)$ for $K$ classes and the minimum is 0. The decision tree, therefore, aims to minimise the GI and reduce its value further at each subsequent node. The GI was used in the Decision Tree (DT) model as it is better than the normal misclassification rate because it is more sensitive to the node probabilities [8].

Another benefit of using a decision tree over an SVM (or a more complex deep learning algorithm like a CNN), is that the output model can be interpreted more easily. The decision tree is colloquially known as more of a 'glass box' than other deep learning algorithms which are often called 'black boxes', i.e. it is harder to understand what is going on under the hood.

## 3. DATA ACQUISITION

As is common in analysis techniques, the categorical data available to solve the problem will be referred to as 'features'. The following subsection describes the features extracted from an HEVC encoder. For 17 standard JCT-VC test sequences listed in Table 1; the total pool of CU samples on which the algorithms were trained is approximately 4 million.

Information about the motion, texture and prediction error in a frame can be gathered from the inter-prediction process at the encoder. These are often factors in the encoder's selection of CU size, therefore features extracted from this process are a good source for the CU splitting problem, as this happens first and correlations between the two can be expected.

The motion information of a CU, or more precisely a set of adjacent Prediction Units (PUs), is predicted using a similar Lagrangian minimisation technique as described by Eq. (1).

| Class | Resolution | Sequence Names |
|---|---|---|
| A | 2560 × 1600 | Traffic, PeopleOnStreet, Nebuta, SteamLocomotive |
| B | 1920 × 1080 | Kimono, ParkScene, Cactus, BasketballDrive, BQTerrace |
| C | 832 × 480 | BasketBallDrill, BQMall, PartyScene, RaceHorses |
| D | 416 × 240 | BasketballPass, BQSquare, BlowingBubbles, RaceHorses |

**Tab. 1.** The names, class and resolution of the test sequences from the JCT-VC [9] used in this research.

Before this, in a typical HEVC implementation, an initial test is carried out to see if this can be skipped. If so, the motion information from a neighbouring block can be 'merged' to the current block. The merge mode is useful for static images for background areas of constant motion, where motion vectors of adjacent blocks are often highly correlated. Since the testing process of the merge/skip mode only applies to encoding a CU as whole and typically requires much lower computational complexity than brute force sub-block partitioning and motion estimation, it is performed *before* other more complex operations in an encoder. Using this, and information about the current CU as a whole, the features were extracted for the decision tree algorithm. Extracted features include:

- The Rate-Distortion Cost (RDC) from the initial skip/merge test, Figure 3. If the RDC of a CU is higher, there is a higher probability of that CU being split into a smaller depth, making it is a useful feature.
- Bits: The number of bits from the skip/merge test. The distribution is similar to the RDC, as the two features are correlated by the RDO equation.
- Coded Block Flag (CBF): The boolean value is also set in the initial skip/merge test. It indicates whether the block in question has any non-zero transform coefficients. For the purposes of this research, only the luma CBF was considered.
- Average Neighbour Depth (AND): The CUs above and to the left of the current CU will have already been encoded (excluding edge and corner cases). The depth at which these were split can be averaged and provided as the AND. This is especially useful for depth zero in the case of homogeneous areas, where a large area of CUs can be left whole. For smaller CUs the inverse is true i.e. smaller CUs tend to have smaller neighbours, making it a useful feature.
- Skip Flag (SF): Boolean flag indicating whether any residual information is sent on the bit stream for the CU or not. If the skip flag is 'true', the only information sent to the decoder for this CU is the merge mode index, specifying which motion vector from the neighbouring PUs should be used to build a prediction. If the SF is 'true', the CBF is consequently set to 'false'.
- Prediction Mode (PM): PMs define the type of partitions (2N×2N, 2N×N or N×2N) used to compute motion information e.g. motion vectors and reference frame indices. This only applies to the encoding of a CU as a whole without the merge mode, as the merge mode can only be used for 2N×2N partitions.
- Quantisation Parameter (QP) and QP Offset (QPO): Quantisation maps the signal amplitudes to a predefined set of values, optimised to minimise the amount of signal loss. In this research, sequences were encoded using base QPs of 22, 27, 32 and 37.

In HEVC, the QPO for each frame in a group of pictures can have values between 1 and 4 relative to the QP of the leading intra coded frame. The offset depends on the temporal and spatial properties of the specific frame [10]. Frames which are more frequently referenced are given a lower QP whereas those which are seldom referenced can get away with coarser quantisation and thus a higher QP to save bit rate.

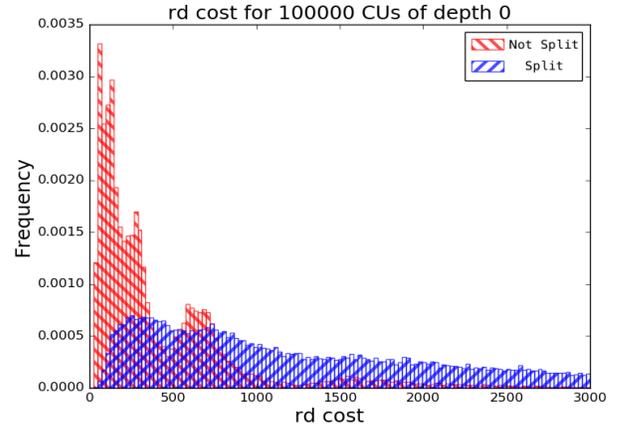

**Fig. 3**. R-D Cost for 100,000 depth 0 (64×64 pixels) coding units, separated into those split and not-split.

| Depth | SF | CBF | RDC | Bits | AND | QP | λ | QPO | PM |
|---|---|---|---|---|---|---|---|---|---|
| 0 | 0.57 | 0.50 | 0.38 | 0.28 | 0.40 | 0.22 | 0.19 | 0.17 | 0.41 |
| 1 | 0.58 | 0.53 | 0.54 | 0.50 | 0.58 | 0.21 | 0.19 | 0.17 | 0.54 |
| 2 | 0.45 | 0.39 | 0.56 | 0.57 | 0.42 | 0.13 | 0.12 | 0.11 | 0.48 |

**Tab. 2**. Absolute values of the feature correlations from 100,000 CUs with the 'truth' split decision.

Table 2 shows the Pearson correlations of the features with the true split decision for 100,000 depth 0 (64×64) coding units chosen at random from Classes A to D.

## 4. EXPERIMENTATION AND DEVELOPMENT

Following feature extraction and correlation analysis, the decision tree models could be created and tested. The number of nodes in a DT model increase as $2^n$ where $n$ starts at 0 and is referred to as the 'depth' of the tree. With no maximum depth limit (or other limits in place), the tree will continue to grow until all the nodes are 'pure', meaning that each leaf only contains samples belonging to a single class. The maximum tree depth was set at 5. This comes with the sacrifice that some leaves are left impure, but for large datasets, the tree can quickly become very complex which this research aimed to avoid.

To further reduce complexity, the minimum number of samples allowed in a leaf can be assigned. Allowing for coarser classification which reduces the likelihood of classifying outliers and therefore the likelihood of the model becoming over-trained. In this research, the minimum number of samples allowed per leaf was set to 0.1% of the total sample size. In most tests, 100,000 samples were used, therefore giving a minimum leaf size of 100. Continuous data was also binned where applicable to avoid the node split criteria becoming too specific.

A novel tool for 'manual pruning' was developed, motivated by the desire to keep the model as simple as possible. Even though the decision tree is one of the simplest 'machine learning' algorithms, importing all the criteria from a DT model trained on many thousands of samples can become complex. Therefore, for simplicity the tool was written such that two threshold values can be set by the user. These thresholds would discard parts of the tree which did not meet the specified criteria and keep the optimal nodes, resulting in only the simplest and most effective model remaining.

One threshold indicates the percentage accuracy of the decision criteria, and the other indicates the percentage of samples classified by this criterion. In an ideal scenario, a solution could be found which covers a high number of input samples and is also highly accurate (few misclassifications). To help set the thresholds, a plot such as Figure 4 is produced. Nodes which have a peak in the percentage of samples classified (blue line) while the accuracy percentage line (red) remains high are useful classifiers, as one simple statement can be responsible for classifying a large chunk of the samples with satisfactory accuracy.

The threshold values mean that only criteria at nodes which are above both thresholds are output for the user to include in the codec (orange circles), if none are output, the thresholds must be lowered. The allowance of the user to manually set thresholds to remove or 'prune' nodes from the tree which do not fall above the desired threshold is a technique for reducing the complexity of the machine learning model. In the case of this research, if a sample did not fall in the criteria covered by the chosen nodes, the default RDO process was performed as before.

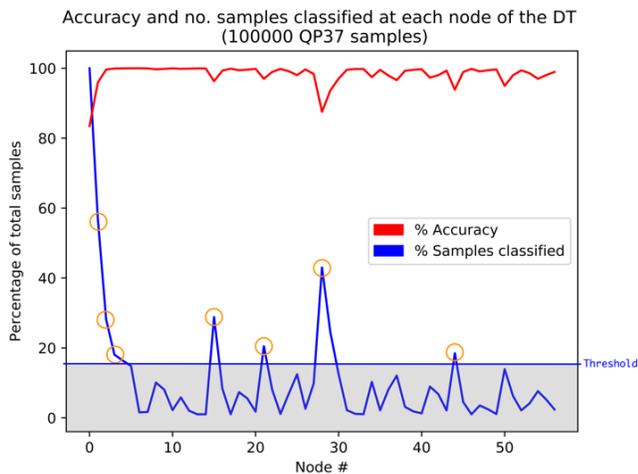

**Fig. 4**. An example line graph produced by the decision tree tool as a threshold setting aid. Here, only nodes which classify at least 17% of samples are included.

## 5. RESULTS

Each of the JCT-VC test sequences were encoded four times with QPs of 22, 27, 32 and 37 to collect the feature set described in Section 3. $k$-fold cross validation was used where $k = 5$ to combat over-training. The thresholds were set to optimise for accuracy, therefore any node whose decision criteria provided an accuracy of less than 97% was discarded/pruned. Once the summary of node criteria with >97% accuracy had been output, the ones which independently classified the largest percentage of samples for each depth were chosen for implementation in the codec.

Three decision tree models were created, one for each 'depth' as per the quad-tree partition structure shown in Figure 2. For simplicity, only one criteria for each depth was chosen, see Table 3 for an example. The criteria were written into the codec using simple statements inserted directly after testing the merge/skip mode and the encoding of the current CU as whole, in order to extract the features needed for the decision criteria. If the statements yielded 'true' then further RDO is skipped and the split flag is set to 0, i.e 'do not split', without performing the more complex recursive splitting encoding tests.

| Depth | Skip RDO Criteria |
| --- | --- |
| 0 | Bits < 50, PM = 0 & RCD < 145 |
| 1 | PM, CBF = 0 & AND < 1.75 |
| 2 | Bits < 50 & PM = 0 |

**Tab. 3**. Best criteria determined by the decision tree.

The practical implementation of HEVC used in this research, the Turing Codec [11], was first used to encode the sequences without the addition of the DT-learned criteria. Features were extracted from some of these (and some were left for testing). The sequences were then encoded a second time with Turing Codec plus the addition of the newly added skip criteria.

The results in Table 4 show that for an average luma detriment BD rate of 0.7%, the encoding time was reduced by 42.1% compared with the time taken by the Turing codec anchor, by the addition of the metric determined by the DTs. In other words, the encoder has been gained a speed up by a factor of 1.73×, by skipping the RDO process for CUs which were deemed to be non-split with a high probability.

| Class | Y BD Rate | U BD Rate | V BD Rate | Enc. Time Δ |
| --- | --- | --- | --- | --- |
| A | 0.6% | -1.0% | -1.2% | -41.7% |
| B | 0.8% | -1.3% | -1.4% | -49.5% |
| C | 0.9% | -1.4% | -0.3% | -39.6% |
| D | 0.7% | -0.9% | -0.9% | -37.6% |
| Avg. | 0.7% | -0.9% | -1.0% | -42.1% |

**Tab. 4**. Encoder performance comparison between the practical implementation codec (anchor) and the anchor with the decision tree metric incorporated, where Y is luma and U and V are chroma.

## 6. CONCLUSIONS

In conclusion, when applying machine learning techniques to video coding problems, complexity of the classification model was found to be an important factor. Due to the rapid ingestion of coding unit data, the same operation is often performed thousands of times as the samples are fed through the encoder. As a result, if these processes are not optimised, the speed of the algorithm can be drastically increased.

The method described in this paper, provided an average encoding time decrease for the practical implementation of HEVC of 42.1%, for a Y BD rate increase of 0.7%. The success was largely attributed to the simplicity of the implementation of the model, which can be written as a collection of cuts on the data.

The manual setting of the accuracy and coverage thresholds make this method flexible and appropriate to tackle many problems in video coding. In the case of this research it was applied to RDO skipping for inter predicted CUs, but it could be applied anywhere in an encoder where features are available. Following this research, initial tests in other video coding frameworks have also indicated positive results.